\documentclass[draft]{agujournal2019}
\usepackage{url} %this package should fix any errors with URLs in refs.
\usepackage{lineno}
\usepackage{amsmath}
\usepackage[inline]{trackchanges} %for better track changes. finalnew option will compile document with changes incorporated.
\usepackage{soul}
\linenumbers
\draftfalse

\journalname{Geophysical Research Letters}

\begin{document}

\nolinenumbers
%% ------------------------------------------------------------------------ %%
%  Title
%
% (A title should be specific, informative, and brief. Use
% abbreviations only if they are defined in the abstract. Titles that
% start with general keywords then specific terms are optimized in
% searches)
%
%% ------------------------------------------------------------------------ %%

% Example: \title{This is a test title}

\title{Explainable Offline-Online Training of Neural Networks for Parameterizations: A 1D Gravity Wave-QBO Testbed in the Small-data Regime}

% Offline-Online Training of Neural Networks for Parameterization: Gravity Waves in a 1D-QBO Model as Testbed

%% ------------------------------------------------------------------------ %%
%
%  AUTHORS AND AFFILIATIONS
%
%% ------------------------------------------------------------------------ %%

% Authors are individuals who have significantly contributed to the
% research and preparation of the article. Group authors are allowed, if
% each author in the group is separately identified in an appendix.)

% List authors by first name or initial followed by last name and
% separated by commas. Use \affil{} to number affiliations, and
% \thanks{} for author notes.
% Additional author notes should be indicated with \thanks{} (for
% example, for current addresses).

% Example: \authors{A. B. Author\affil{1}\thanks{Current address, Antartica}, B. C. Author\affil{2,3}, and D. E.
% Author\affil{3,4}\thanks{Also funded by Monsanto.}}

\authors{Hamid A. Pahlavan\affil{1}, Pedram Hassanzadeh\affil{1}, M. Joan Alexander\affil{2}}

\affiliation{1}{Rice University, Houston, TX, USA}
\affiliation{2}{NorthWest Research Associates, Boulder, CO, USA}

%(repeat as many times as is necessary)

%% Corresponding Author:
% Corresponding author mailing address and e-mail address:

% (include name and email addresses of the corresponding author.  More
% than one corresponding author is allowed in this LaTeX file and for
% publication; but only one corresponding author is allowed in our
% editorial system.)

% Example: \correspondingauthor{First and Last Name}{email@address.edu}

\correspondingauthor{Hamid A. Pahlavan}{pahlavan@rice.edu}

%% Keypoints, final entry on title page.

%  List up to three key points (at least one is required)
%  Key Points summarize the main points and conclusions of the article
%  Each must be 140 characters or fewer with no special characters or punctuation and must be complete sentences

% Example:
% \begin{keypoints}
% \item	List up to three key points (at least one is required)
% \item	Key Points summarize the main points and conclusions of the article
% \item	Each must be 140 characters or fewer with no special characters or punctuation and must be complete sentences
% \end{keypoints}

\begin{keypoints}
\item 1D model of quasi-biennial oscillation (QBO) and gravity waves is used as testbed for training neural network (NN)-based parameterizations
\item Offline training NNs in small-data regimes yields unstable QBOs that are rectified by online re-training using only time-averaged statistics
\item Fourier analysis of NNs reveals that they learn specific filters that are consistent with the dynamics of wave propagation and dissipation

\end{keypoints}

%% ------------------------------------------------------------------------ %%
%
%  ABSTRACT and PLAIN LANGUAGE SUMMARY
%
% A good Abstract will begin with a short description of the problem
% being addressed, briefly describe the new data or analyses, then
% briefly states the main conclusion(s) and how they are supported and
% uncertainties.

% The Plain Language Summary should be written for a broad audience,
% including journalists and the science-interested public, that will not have 
% a background in your field.
%
% A Plain Language Summary is required in GRL, JGR: Planets, JGR: Biogeosciences,
% JGR: Oceans, G-Cubed, Reviews of Geophysics, and JAMES.
% see http://sharingscience.agu.org/creating-plain-language-summary/)
%
%% ------------------------------------------------------------------------ %%

%% \begin{abstract} starts the second page

\begin{abstract}

There are different strategies for training neural networks (NNs) as subgrid-scale parameterizations. Here, we use a 1D model of the quasi-biennial oscillation (QBO) and gravity wave (GW) parameterizations as testbeds. A 12-layer convolutional NN that predicts GW forcings for given wind profiles, when trained offline in a \emph{big-data} regime (100-years), produces realistic QBOs once coupled to the 1D model. In contrast, offline training of this NN in a \emph{small-data} regime (18-months) yields unrealistic QBOs. However, online re-training of just two layers of this NN using ensemble Kalman inversion and only time-averaged QBO statistics leads to parameterizations that yield realistic QBOs. Fourier analysis of these three NNs’ kernels suggests why/how re-training works and reveals that these NNs primarily learn low-pass, high-pass, and a combination of band-pass filters, consistent with the importance of both local and non-local dynamics in GW propagation/dissipation. These findings/strategies apply to data-driven parameterizations of other climate processes generally.

\end{abstract}

\section*{Plain Language Summary}

Due to computational limits, climate models estimate (i.e., parameterize) small-scale physical processes, such as atmospheric gravity waves (GWs), since they occur on scales smaller than the models' grid size. Recently, machine learning techniques, especially neural networks (NNs), have emerged as promising tools for learning these parameterizations from data. Offline and online learning are among the main strategies for training these NN-based parameterizations. Offline learning, while straightforward, requires extensive, high-quality data of the small-scale processes, which are scarce. Alternatively, online learning only needs time or space-averaged data based on large-scale processes, which are more accessible. However, online learning can be computationally expensive. Here, we explore various learning strategies using an NN-based GW parameterization, within a simple model of the quasi-biennial oscillation (QBO), an important quasi-periodic wind pattern in the tropics. When supplied with a large 100-year dataset, the offline-trained NN accurately replicates wind behaviors once coupled to the QBO model. Yet, when limited to an 18-month training dataset (which is more realistic), its performance degrades. Interestingly, by online re-training specific parts of this NN using only time-averaged QBO statistic, its accuracy is restored. We term this approach an ``offline-online" learning strategy. Our findings also benefit parameterization efforts for other climate processes.

%% ------------------------------------------------------------------------ %%
%
%  TEXT
%
%% ------------------------------------------------------------------------ %%

%%% Suggested section heads:
\section{Introduction}
%and even numerical weather prediction models
Due to the current resolution limitations of general circulation models (GCMs), many crucial subgrid-scale (SGS) physical processes remain unresolved, and are instead represented through parameterization. The conventional physics-based parameterizations are based on simplified theories, which introduce significant uncertainties in climate modeling. Recently, machine learning (ML) techniques, particularly deep neural networks, have emerged as novel tools to develop parameterizations. Different strategies exist for training these ML-based parameterizations. In the common offline learning approach, the learnable parameters of neural networks (i.e., weights and biases) are trained using stochastic gradient descent through backpropagation to find a nonlinear mapping between the resolved and SGS processes. However, this approach demands an extensive training dataset that includes the true SGS terms obtained from high-fidelity sources such as high-resolution observations and/or simulations, which are typically scarce. To add to the challenge, the true SGS terms must be properly extracted from these sources, which can be sensitive to separation methods, as well as the filtering and coarse-graining operations \cite{grooms2021diffusion, sun2023quantifying}. 

% Additionally, even when ML-based parameterizations perform well offline, they can still lead to instabilities during inference \cite{bretherton2022correcting}.

%, that is, when coupled into GCM simulations
%addresses many prior issues by allowing
%this method's primary challenge is that 

Alternatively, learning SGS parameterization can be approached as an ``online task'', which allows learning from partial observations or statistics. However, learning from statistics requires performing long-term simulations during training, making the learning process challenging \cite{schneider2023harnessing}. For online learning, various methods such as reinforcement learning \cite{novati2021automating}, differentiable programming \cite{frezat2022posteriori, gelbrecht2023differentiable}, and ensemble Kalman inversion (EKI) \cite{iglesias2013ensemble, lopez2022training} can be used. Here, we use EKI, a gradient-free algorithm~\cite{kovachki2019ensemble}, which is ideal for GCMs where computing derivatives can be challenging. 

% However, initializing EKI can be complex, with poor initializations potentially causing instabilities and learning failures.

%exerting a substantial influence on the dynamics of the middle and upper atmosphere 
%, as outlined in \citeA{kruse2023gravity} and the accompanying references
%GWs play a crucial role in the transport of momentum through Earth's atmosphere \cite{fritts2003gravity}.
%GWs are disturbances with horizontal wavelengths ranging from tens to thousands of kilometers. 

Atmospheric gravity waves (GWs) are among the physical processes that are not fully resolved in the current GCMs as they span scales of $O(1)$ to $O(1000)$~km. GWs play a crucial role in the transport of energy and momentum through the atmosphere \cite{fritts2003gravity}. With decades of developments, GW parameterization (GWP) is now a critical component of GCMs for reproducing realistic atmospheric circulation mean and variability \cite{kruse2023gravity}. For instance, GCMs require skillful GWPs to naturally produce the quasi-biennial oscillation (QBO) \cite{richter2020progress}, which is characterized by the downward propagation of successive westerly and easterly winds with an average period of $\sim$28 months \cite{baldwin2001quasi}. The QBO is the primary mode of interannual variability in the tropical stratosphere with links to subseasonal-to-seasonal forecast skills \cite{anstey2022impacts}. GWs are believed to contribute significantly to the forcing of the QBO \cite{kawatani2010roles, ern2014interaction, richter2014simulation, kim2015contributions, pahlavan2021revisiting}.

Recently, ML has been used to emulate or calibrate existing physics-based GWP schemes \cite{chantry2021machine, espinosa2022machine,mansfield2022calibration}, or to estimate the GWs variability or structure from high-resolution reanalysis data \cite{matsuoka2020application, amiramjadi2023using}. The much more ambitious and challenging task of learning data-driven GWP schemes and coupling them to GCMs, for example, to improve the simulation of QBO, is the subject of growing efforts and the datasets needed for such work have just started to emerge~\cite{sun2023quantifying}.

In this study, we use a conceptual 1D model of the QBO and GWP as testbeds to explore various learning strategies and challenges arising from scarcity of high-fidelity training data to inform future studies with GCMs. We show that a 12-layer convolutional neural network (CNN)-based GWP, when trained offline in a \emph{big-data} regime spanning 100 years, generates accurate QBOs once coupled to the 1D model. However, offline training the CNN in a \emph{small-data} regime covering only 18 consecutive months yields unrealistic QBOs. This \emph{small-data} regime represents the common situations with limited availability of high-quality SGS data for training. Remarkably, by selectively online re-training only two layers of this CNN with EKI and using only time-averaged QBO statistics, we obtain GWPs that reproduce realistic QBOs. We refer to this approach as ``offline-online" learning. We also use the Spectral Analysis of Regression Kernels and Activations (SpArK) framework (introduced in \citeA{subel2023explaining}) to provide physically interpretable insights into what these three CNNs learn. While this study primarily addresses GWP, the findings are expected to be applicable broadly to data-driven parameterizations of other climate processes.

% , a challenge frequently encountered
% a Fourier analysis of the kernels from these three CNNs reveals that they predominantly learn specific filters, which are consistent with the dynamics of GW propagation. This analysis also offers insights into the mechanisms behind the success of online re-training. 
% \cite{shamir2022gravity}

\section{Methods}

\subsection{1D-QBO Model}

The 1D-QBO model represents a 1D model of the tropical stratosphere \cite{holton1972updated, plumb1977interaction}. With a source of parameterized waves at its lower boundary, the model is a minimal configuration that represents the wave-mean flow interaction. In this study, the 1D-QBO is structured as a forced advection-diffusion model:

\begin{linenomath*}
\begin{equation} \label{eq:1}
\frac{\partial{u}}{\partial{t}} + \omega \frac{\partial{u}}{\partial{z}} - \kappa \frac{\partial^2{u}}{\partial{z}^2} = G(u) + \eta(t)
\end{equation}
\end{linenomath*}

\noindent with zonal wind $u (t, z)$ as a function of time $t$ and height $z$, upwelling $\omega = 0$, diffusivity $\kappa= 0.3 $ m$^2$ s$^{-1}$, and GW drag $G$. By setting $\omega = 0$, we exclude vertical advection for simplicity. $\eta$ is a stochastic forcing, which represents the missing physics within the 1D model, and its importance is further detailed in the Supporting Information S1.

Following \citeA{plumb1977interaction}, the model is driven by two vertically propagating GWs with zonal phase speeds $(c_1, c_2) = (-30, +30)$ m s$^{-1}$. As these waves propagate upward, they dissipate and force the mean flow toward their phase speed. The vertical group velocity of these waves depends non-linearly on the difference between their phase speed and the mean flow, becoming smaller when the zonal wind is close to its phase speed, and reaching zero at a critical level where $u = c$. With constant dissipation, slower ascent results in more dissipation per unit height. 

% The described wave propagation dynamics are visually illustrated
%which is independent of height
%This amplitude gradually decreases closer to the top and bottom boundaries 

Figure \ref{Fig. 1}a shows sample profiles of GW drag (GWD) and zonal wind, illustrating the downward propagation of a QBO westerly phase from the upper boundary. Note the concentration of GWD over a shallow vertical extent where the eastward wave ($c = 30$ m s$^{-1}$) reaches its critical level. At this layer, the wave breaks and transfers its momentum to the mean flow, causing the GWD at higher levels to become zero.

More details on the model configuration are provided in Supporting Information S1. This setup yields an oscillation with a period ($\tau$) of 28.0$\pm$0.7 months, and an amplitude ($\sigma$) of 21$\pm$0.3 m s$^{-1}$ at the 25 km altitude (Figure~\ref{Fig. 1}b). The standard deviations of the period and amplitude are based on $\sim$430 QBO cycles in a 1000-year simulation. This simulation is our ``truth" and is used to evaluate the performance of CNN-based GWP. Using the same setup, we produce an independent 100-year dataset specifically for training and validation purposes.

\subsection{CNN-based GWP}

We explore various learning strategies by emulating the GWD, $G(u)$ in Equation \eqref{eq:1}, using a CNN, denoted as $G_{\mathrm{CNN}}(u, \theta)$, with $\theta$ being the learnable parameters of the CNN. This CNN consists of 12 sequential 1D convolutional layers. Each hidden layer has 15 channels, each with 15 kernels with a size of 5, resulting in $\sim$11,600 learnable parameters. The activation function is hyperbolic tangent (tanh). Training a CNN means to learn the parameters $\theta$, either offline or online, as detailed below.

\subsection{Offline Learning}

Offline learning seeks to find the optimal $\theta$ values by matching $G_{\mathrm{CNN}}$ and the true $G$ profiles for a given profile of $u (t, z)$, which is achieved by minimizing the following loss:

\begin{linenomath*}
\begin{equation} \label{eq:2}
\mathcal{L}\,_{offline} = \frac{1}{n} \sum_{i=1}^{n} \,\,  \bigl\| G(u_i) - G_{\mathrm{CNN}}(u_i, \theta) \bigl\|_2^2
\end{equation}
\end{linenomath*}

\noindent where $n$ is the number of training samples and $\|.\|_2$ is the $L_2$ norm. We train the CNN offline under two distinct data regimes: a) \emph{big-data}, denoted as CNN-BD, which uses 100 years of sequential data, representing an ideal scenario with ample data, and b) \emph{small-data} (CNN-SD), which includes only 18 consecutive months of data, representing a more realistic scenario given the cost associated with, for instance, GW-resolving global simulations. Further insights on using a more strategically sampled 18 months, instead of a continuous span, are detailed in the Discussion section.

% , reflecting scenarios where availability of high-resolution simulations or observations may be limited

\begin{figure}
\noindent\includegraphics[width=\textwidth]{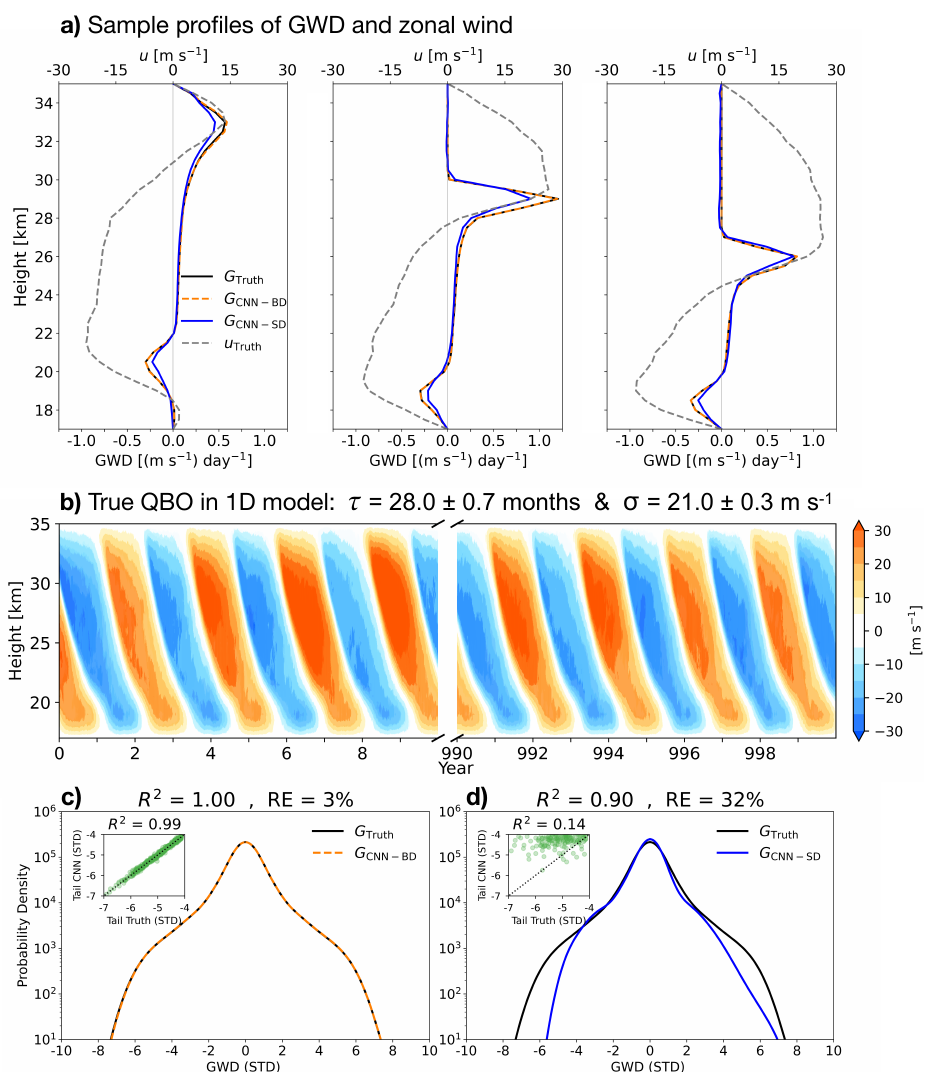}
\caption{(a) Sample profiles of the true GWD ($G$) and zonal wind ($u$), spaced 150 days apart. Also shown is the \emph{a-priori} (offline) predicted GWD from CNNs that predict $G$ as a function of $u$, and are trained either in the \emph{big-data} regime (CNN-BD) or in the \emph{small-data} regime (CNN-SD). (b) Time-height section of zonal wind of the true QBO in the 1D model (see Methods). Note the time axis break, such as only the first and last 10 years of the simulation are shown. The \emph{a-priori} performances of CNN-BD, and CNN-SD are shown in panels (c) and (d). (c) Probability density function (PDF) of the true and predicted $G$ by CNN-BD. (d) As in (c), but for $G$ predicted by CNN-SD. In panels (c), and (d), the insets show scatter plots representing the tails of the PDFs, identified as the top 1\% of magnitudes. The $x$-axis is normalized by the standard deviation (STD). In these panels, $R^2$ is the squared of the Pearson correlation coefficients between true and predicted GWD. Relative error (RE) is defined as $|G-G_{\mathrm{CNN}}|/|G|$, where $|.|$ denotes the average of absolute values over all model levels.}
\label{Fig. 1}
\end{figure}

\subsection{Online Learning}

In online learning, we learn the parameters $\theta$ by using time-averaged statistics of the QBO, and minimizing the following loss:

\begin{linenomath*}
\begin{equation} \label{eq:3}
\mathcal{L}\,_{online} = \bigg\| \mathcal{H}\Bigl(\Psi\bigl(u, G(u)\bigl)\Bigl) - \mathcal{H}\Bigl(\Psi\bigl(u, G_{\mathrm{CNN}}(u, \theta)\bigl)\Bigl) \bigg\|_\Gamma^2 
\end{equation}
\end{linenomath*}

\noindent where $\|.\|_\Gamma$ is the Mahalanobis norm, $\Gamma$ denotes the variance of the system's internal noise, and $\Psi$ is the forward model, the numerical solver of the 1D-QBO model in this case, $\mathcal{H}$ is the observational map, which encapsulates all averaging and post-processing operations necessary to derive the desired statistics from an observable field, zonal wind $u$ in this case. See \citeA{lopez2022training} for more details.

Various optimization methods can be used to minimize $\mathcal{L}_{online}$. As highlighted earlier, we employ EKI, which has been increasingly used for parameter estimation in recent climate studies \cite{cleary2021calibrate, oliver2022ensemblekalmanprocesses, lopez2022training}. Briefly, as an iterative method to solve inverse problems, EKI starts with an ensemble of model parameters $\theta$ drawn from a prior distribution. As the iterations proceed, these parameters are updated based on the discrepancies between statistics simulated with the model and the true statistics, usually obtained from observations, reanalysis, or high-resolution simulations. Once the algorithm converges, the optimal parameter values are then the ensemble mean of the last iteration.

The 1D model is very sensitive to the changes of GWD profiles, and for some sets of parameters, simulations become physically or numerically unstable, preventing the EKI algorithm from converging. We address these model failures following the methodology proposed in \citeA{lopez2022training}. 

% The formulation in Equation \eqref {eq:3} offers a significant advantage by sidestepping the need to match $G_{\mathrm{CNN}}(u, \theta)$ and $G(u)$, which is necessary in offline training. Instead, it leverages statistically aggregated data based on an observable field for parameter estimation. This is particularly beneficial considering the difficulties in observing and quantifying the true GWD from high-resolution simulations and observations \cite{sun2023quantifying}.

% Furthermore, time-averaged statistics are what is relevant and matter most for climate predictions, rather than instantaneous values of subgrid-scale terms.

For the online training of the CNN using EKI, denoted as CNN-EKI, our setup includes 200 ensemble members, 10 iterations, and 85 statistics, derived from a 10-year span of zonal wind $u$ from our true QBO simulation. We run each model for 15 years, then calculate the desired statistics from the last 10 years of those runs. The EKI's efficacy can be notably impacted by these choices, with poor selections potentially causing instabilities, underscoring the challenges associated with online learning. Section 3.2 offers further details on the statistics and prior distributions used in this study.

\section{Results}

\subsection{Offline Learning in the Big and Small-data Regimes}

% As mentioned earlier, our evaluation uses an independent 1000-year simulation, which has not been exposed to the CNN during offline learning, and is referred to as ``truth".

We begin by evaluating the \emph{a-priori} (offline) performances of the CNN-BD and CNN-SD. Figure \ref{Fig. 1}a shows sample GWD profiles predicted by these two CNNs, compared with the true GWD profiles. Both CNNs capture the general structure of the true GWD. However, $G_{\mathrm{CNN-BD}}$ aligns perfectly with the true GWD profiles, while $G_{\mathrm{CNN-SD}}$ exhibits some discrepancies, especially in representing the peaks.

Figure \ref{Fig. 1}c compares the probability density function (PDF) of $G_{\mathrm{CNN-BD}}$ with that of the truth. The outstanding \emph{a-priori} performance is evident from their overlap, further highlighted by a mere 3\% relative error (RE). In contrast, the \emph{a-priori} performance of the CNN-SD, shown in Figure \ref{Fig. 1}d, clearly diminishes, evidenced from the increase in the RE to 32\%. Furthermore, a pronounced decline in $R^2$ can be observed at the tails, decreasing from 0.99 to 0.14. These findings are in line with expectations. When abundant training data are provided, deep NNs such as our CNN can be effectively trained offline and demonstrate accurate \emph{a-priori} performance. Conversely, with limited data, the accuracy decreases, especially when predicting rare (but large) events at the tails. It is noteworthy that in the context of the \emph{small-data} regime, our extensive experiments with smaller CNNs and other NN architectures with fewer parameters did not yield successful results (not shown).

% Next, we will examine the \emph{a-posteriori} (online) performance of the CNN under these two data regimes.
% Consequently, at any given time step, the CNN receives a profile of $u$ and produces a corresponding GWD profile, which forces the subsequent profile of $u$. 
%The instability of CNN-SD emphasizes the potential disconnect between \emph{a-priori} metrics and \emph{a-posteriori} performances, as one might not anticipate such instability from the GWD profile samples in Figure \ref{Fig. 1}a and an $R^2$ of 0.90. However, the PDF in \ref{Fig.1}d more clearly reveals the shortcomings of CNN-SD.
%(Figure~\ref{Fig. 1}b) and it remains stable over a 1000-year span
%particularly at the upper levels, 
%The latter exhibits pronounced sensitivity to changes in GWD.

\begin{figure}[t]
\noindent\includegraphics[width=\textwidth]{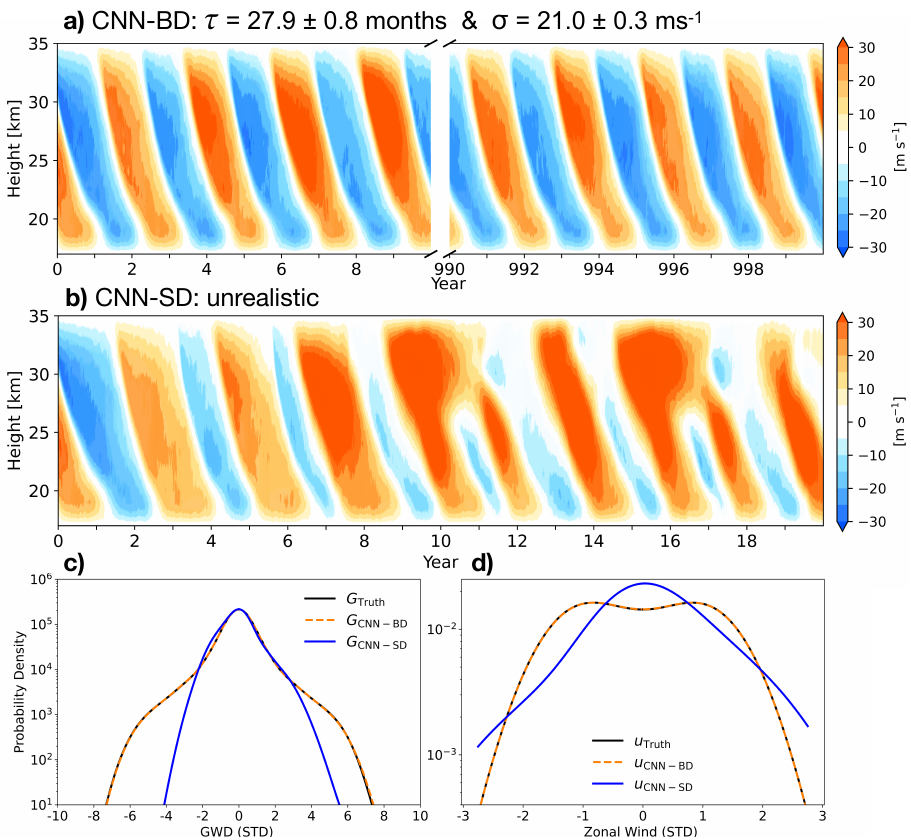}
\caption{ (a) The \emph{a-posteriori} (online) performance of the CNN-BD. The QBO remains stable for 1000 years, with its period and amplitude closely matching the true QBO. (b) As in (a), but for the CNN-SD. The QBO is unrealistic. (c) PDFs of the true and \emph{a-posteriori} predicted GWD ($G$). (d) As in (c), but for zonal wind ($u$).}
\label{Fig. 2}
\end{figure}

In the \emph{a-posteriori} (online) evaluations, we replace $G(u)$ with $G_{\mathrm{CNN}}(u, \theta)$ and run the model for 1000 years. Figure~\ref{Fig. 2}a shows the \emph{a-posteriori} performance of the CNN-BD, demonstrating a QBO whose structure, period, and amplitude closely match that of the true QBO. In contrast, Figure~\ref{Fig. 2}b reveals that the QBO simulated with the CNN-SD becomes unrealistic after the initial QBO cycles, with intensified westerly phases and diminished easterly phases. The PDFs of GWD and zonal wind are presented in Figures~\ref{Fig. 2}c and~\ref{Fig. 2}d. The indistinguishable overlap between the PDFs of truth and the CNN-BD underscores its outstanding \emph{a-posteriori} performance. In contrast, the PDFs for the CNN-SD demonstrate a significant deviation from those of the truth.

This specific unrealistic behavior of the CNN-SD is a result of the specific 18-month segment used for training. When we choose a different 18-month segment, the QBO exhibits other unrealistic deviations. The key takeaway, however, is that an 18-month sequential dataset is not adequate to achieve accurate \emph{a-posteriori} QBO in the 1D model. Notably, \citeA{espinosa2022machine} achieved stable \emph{a-posteriori} QBO by training their ML-based emulator of the physics-based GWP using only 12 months of data, which were dominated by the westerly phase of the QBO. However, their use of global data suggests that the emulator might have learnt from regions with easterly winds outside the tropical stratosphere.

%Furthermore, it is important to note that the QBO is considerably more constrained in GCMs compared to the 1D model.

% In practice, this corresponds with availability of extensive GW-resolving simulations, as provided by libraries like \citeA{sun2023quantifying}, which can offer numerous, relatively independent profiles of GWD.

Our results so far indicate that the CNN-BD shows outstanding \emph{a-priori} and \emph{a-posteriori} performances. In contrast, the CNN-SD yields unstable \emph{a-posteriori} QBO and, given the appropriate metrics (e.g., PDF tails), a poor \emph{a-priori} performance too. This finding prompts one of the central questions of this study: Is it possible to use online learning to improve the CNN-SD and rectify this unrealistic QBO behavior? Note that the terms \emph{small-} and \emph{big-data} regimes in our context refer to the number of $G$ snapshots available for offline learning. For online learning in the \emph{small-data} regime, we assume that we have access to the time-averaged statistics of the true QBO but have limited snapshots of $G$. 

In the context of online learning, one approach might be to train a CNN from scratch (i.e., from random initialization of $\theta$) using methods like EKI, and the QBO statistics as the targets. However, as we discuss below, the priors (initialization of $\theta$) are critical for convergence of online learning methods, and poor priors, such as random ones, can lead to failed learning. Another approach, which we pursue here, is to use the parameters of the CNN-SD as priors. We will refer to this as the ``offline-online" learning approach. Note that this approach basically performs transfer learning \cite{chattopadhyay2020data,subel2021data, guan2022stable, subel2023explaining}, and in fact, similar to transfer learning, it requires re-training only a few CNN layers, as further elaborated below.

% Selecting these statistics may not always be straightforward and often depends on the physics of the system.
%characteristics, particularly its 

\subsection{Offline-online Learning in the Small-data Regime}

A major question in using statistics for parameter estimation, as in the EKI method, is determining whether the targeted statistics are adequate to constrain the learnable parameters, which could be high-dimensional. Here, we began by using the common targets for QBO, the period and amplitude, as our time-averaged statistics. However, we found that various unrealistic oscillations can misleadingly mimic the true QBO's period and amplitude, including an upward propagating QBO, suggesting non-uniqueness of the parameters and under-constrained optimization. To address this, we expanded our targeted statistics to include cross-covariances between various QBO levels, aiming to better capture its downward propagation. This led us to use an extensive list of 85 statistics (see the Supporting Information S1).

% If the QBO was more constrained in the 1D model, as is the case for the QBO in GCMs, we likely could have used a smaller set of statistics for parameter estimation.

 % considered for each phase of the QBO separately, 
 %, calculated as standard deviation of zonal wind at several levels, 
 % and broadly in any Bayesian analysis
 % For instance,  used EKI to online calibrate two key parameters of a physics-based GWP scheme in an idealized moist atmospheric model. They used log-normal priors knowing that these parameters, GW stress and phase speed, must always be positive. They also used domain knowledge to inform the choice of the mean and variance of these prior distributions. 

 %When learning parameters of a physics-based parameterization, one can often lean on prior knowledge of the system's physics and impose constraints on the prior distributions, for instance as seen in \citeA{mansfield2022calibration}. However, such a prior knowledge is often not available when learning parameters for an ML-based parameterization.
  % Their role is crucial as they can significantly influence the posterior distribution and consequently the optimal values of the parameters
  % It is noteworthy that \citeA{subel2023explaining} offers more structured strategies for determining the optimal layers for re-training in the context of transfer learning.
 
As discussed above, another key element of EKI, and other online learning methods, is the prior distributions. We use the weights and biases from CNN-SD as our priors. These serve as the mean values for unconstrained Gaussian priors with a standard deviation of $0.01$, given that their magnitude is around $O(10^{-1})$. Through further trial-and-error experiments, we discovered that it is unnecessary to online re-train every layer of the CNN-SD to achieve a stable QBO. By online re-training of only the shallowest and deepest hidden layers of the CNN-SD (i.e., layers 2 and 11), we obtain results that are on par with full re-training. Consequently, we confine our subsequent discussions to these findings. By re-training only two layers, we engage a significantly reduced parameter set, simplifying the analysis and enhancing interpretability, which we discuss in the next section.

% fine-tune the parameters, 

During the online re-training process, the EKI error decreases sharply in the first iteration (See also Figure S2 in Supporting Information S1). Subsequent iterations further reduce the mismatch between the statistics of the model (1D-QBO with CNN-based GWP) and the true QBO statistics. The ensemble mean of the parameters from the last iteration is then considered as the optimal parameter set for the CNN, referred to as CNN-EKI.

\begin{figure}[t]
\noindent\includegraphics[width=\textwidth]{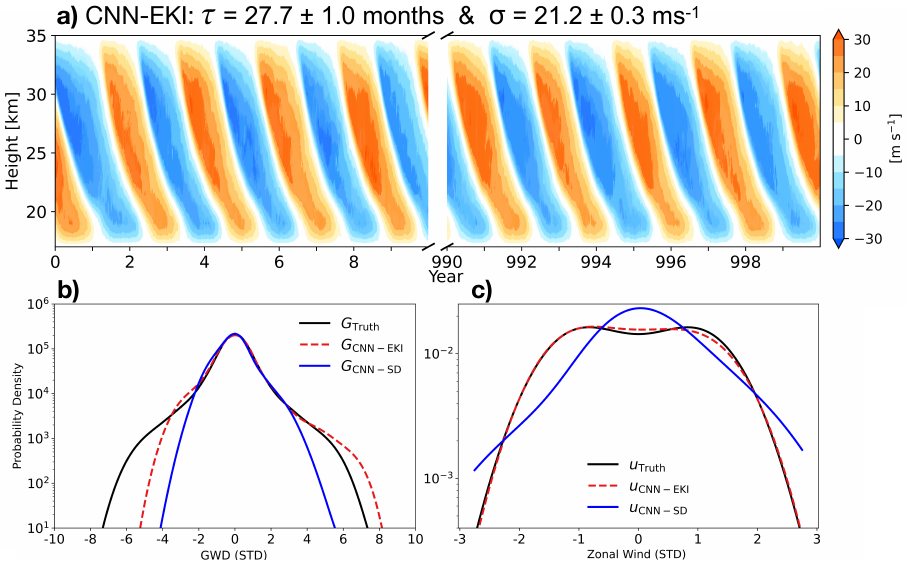}
\caption{(a) The \emph{a-posteriori} performance of the CNN after online re-training the CNN-SD, referred to as CNN-EKI. (b) PDFs of the true and \emph{a-posteriori} predicted GWD ($G$), before and after online re-training. (c) As in (b), but for zonal wind ($u$).}
\label{Fig. 3}
\end{figure}

The \emph{a-posteriori} performance of the CNN-EKI is illustrated in Figure~\ref{Fig. 3}. Panel (a) shows an accurate QBO with period and amplitude closely agreeing with those of the true QBO. Figures~\ref{Fig. 3}b and c show the PDFs of GWD and zonal wind before and after the online re-training, with a comparison to the truth. The zonal wind's PDF demonstrates a significant improvement, closely aligning with the true QBO, albeit with minor deviations. This is despite the smaller improvement in the PDF of CNN-EKI's GWD, which better matches the PDF of the truth, but still deviates at the tails beyond four standard deviations. This could be expected, considering that we only use the statistics of zonal wind for online re-training, which does not necessarily constrain the PDF of GWD. In other words, the values of GWD beyond four standard deviations, which occur orders of magnitude less frequently than GWD values within two standard deviations, do not heavily influence the QBO period, amplitude, and its overall structure. Also, note that improvement in \emph{a-posteriori} performance without any improvement to the \emph{a-priori} one (or even its degradation) is a common feature of online learning, as reported in several past studies \cite{gelbrecht2023differentiable}.

% Consistent with the \emph{a-posteriori} results of panel (a), t

% 

\subsection{Explainable Learning using SpArK}

%in offline and online learning

Next we use the SpArK framework \cite{subel2023explaining} to gain insights into the inner workings of the CNNs and connect them to the underlying physics of the GW propagation. Briefly, \citeA{subel2023explaining} applied Fourier transformation and convolution theorem to the governing equations of CNNs. They showed that the kernels of CNNs used for SGS closure modeling of turbulent flows, while seem meaningless in the physical space, are meaningful spectral filters in the Fourier domain, comprising low-, high-, band-pass, and Gabor filters.

%They further demonstrated that examining changes in the spectra of the kernels before and after re-training can explain the physics learned during transfer learning.
%equivalent to the total number of vertical levels
%across various layers
% Though presenting all 15$^2$ kernels per layer would be excessive,

%Similar results can be achieved by clustering the kernels using the $k$-means algorithm following \citeA{subel2023explaining}, although these are not shown here.

We start by extending each CNN kernel, originally of size 5 (doing convolution on activations of size 37), to match the size of the activations by zero-padding, resulting in kernels of size 37 (activation is the output of a layer after applying filters and non-linearity). We then simply transfer them into a spectral space using a Fourier transform. Upon close inspection of the Fourier spectra of kernels, it becomes evident that they are a combination of low-, high-, and band-pass filters. The similarity across the spectra of many kernels allows us to meaningfully categorize them by their dominant wavenumber ($k^*$), that is, the wavenumber where the spectrum peaks in magnitude.

% , with black lines denoting kernels trained in the \emph{big-data} regime.

% Notably, when training the CNN-BD from a different random initialization, the results remain consistent: the same wavenumbers dominate, albeit with slight variations in their frequencies.

%as is evident from the integral operator in Equation 1 in the Supporting Information S1
%whether physics-based or ML-based

\begin{figure}[t]
\noindent\includegraphics[width=\textwidth]{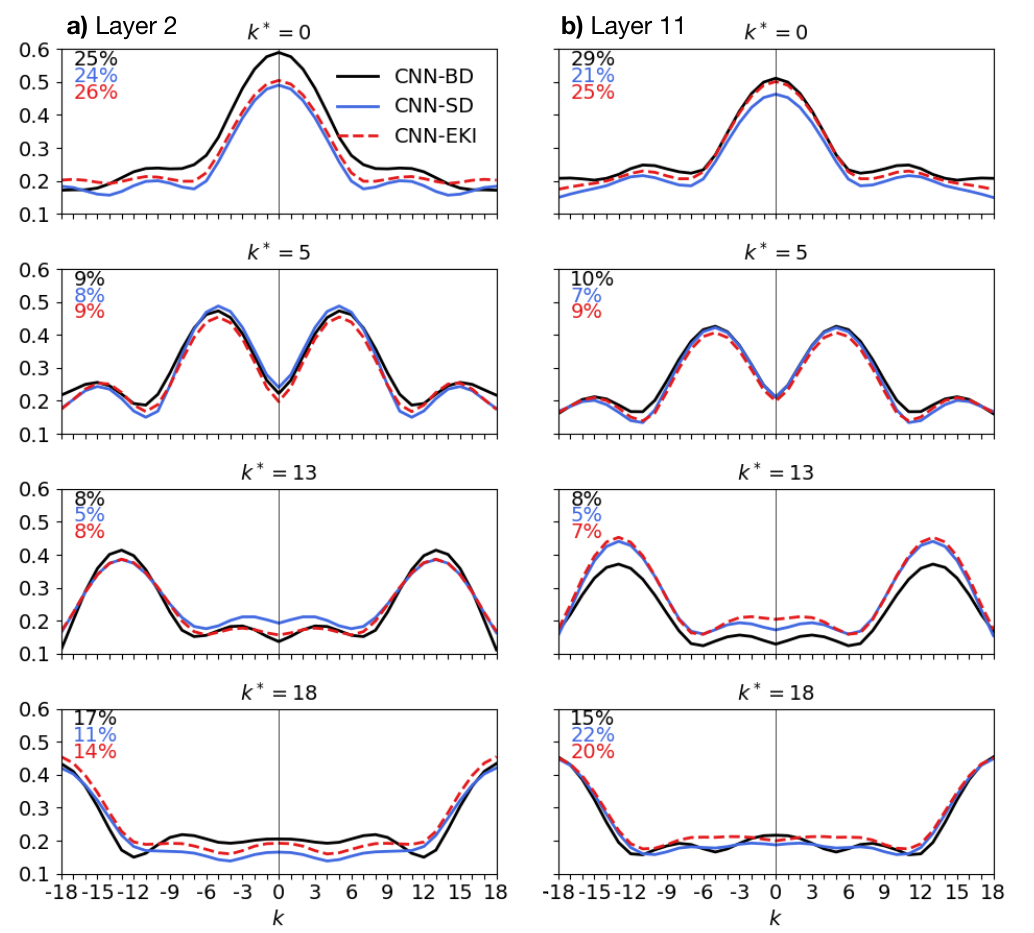}
\caption{The averaged Fourier spectra of kernels for the four most frequent wavenumber peaks $k^*$ for (a) layer 2 and, (b) layer 11 for the three CNNs. The frequency of each kernel within its respective layer and for each of the three CNNs is indicated in the top left corner of each panel.}
\label{Fig. 4}
\end{figure}

The averaged spectra of kernels for the four most frequent wavenumbers are shown in Figure \ref{Fig. 4}. The dominant spectra have $k^*=0$ (low-pass filters), followed by $k^*=18$ (high-pass filters). $k^*=5$ and $k^*=13$ come next, each representing band-pass filters. While Figure~\ref{Fig. 4} showcases the composited kernels from layers 2 and 11, similar patterns are observed across other layers. Collectively, these four wavenumbers account for $\sim$65\% of all the kernels, with $k^*=0$ and $k^*=18$ together constituting $\sim$45\% of the total. The frequent appearance of low- and high-pass filters, and to some degree, band-pass filters, can be connected to the dynamics of GW propagation and dissipation. On one hand, the GWD at a given level depends on the local zonal wind conditions. As discussed earlier, a wave propagates upward more slowly when the local wind is close to its phase speed, leading to increased dissipation. This dissipation is especially pronounced near the critical level, highlighting the essential role of local dynamics. On the other hand, the cumulative wind profile below a given level significantly impacts the GWD, underscoring the relevance of non-local dynamics. For any GWP scheme, capturing both local and non-local dynamics is necessary to be able to generate a spontaneous QBO \cite{campbell2005constraints}.

The prevalence of low-pass filters aligns well with the need to capture non-local dynamics, as these filters extract large scales and perform averaging. On the other hand, high-pass filters capture more local dynamics by extracting smaller scales. The band-pass filters, which resemble wavelets, extract specific scales, local in space. That said, the presence of layers and non-linearity further influence the output of each kernel, obscuring further understanding of the role of each kernel in connecting the profiles of the input (zonal winds) to the output (GWDs). Still, the understanding that emerges from SpArK can enable future work to exploit some of the novel mathematical tools from the deep learning community, in particular, those that leverage wavelet-analysis of NNs \cite{mallat2016understanding, ha2021adaptive}. Finally, it should be highlighted that in this study, we focus on the Fourier spectra, and categorize the kernels based on $k^*$. A deeper analysis of both real and imaginary parts of the Fourier transformation of kernels and activations, as well as metrics beyond frequency, will be needed to gain further insight into how the NNs are representing the GW dynamics. 

While the full explainability of each NN remains a challenge, SpArK offers more insight into how a NN changes after online re-training. As shown in Figure~\ref{Fig. 4}, while these four wavenumbers retain their dominance in CNN-SD, their frequency deviates from those observed in CNN-BD. Yet, following online re-training, the frequency of these kernels aligns more closely with those in CNN-BD. This suggests a transformation of kernels from potentially ineffective wavenumbers to more efficient filters. This kind of analysis provides insights into the calibration processes of ML-based parameterizations, ensuring that the adjustments are not only effective but also physically interpretable. 

%We remind the reader that only layers 2 and 11 undergo online re-training, meaning that the kernels in other layers of CNN-SD remain unchanged during this re-training phase.

\section{Discussion on Strategic Sampling and Offline-online feasibility}

We find that a consecutive 18-month span is inadequate for offline training a CNN-based GWP in this 1D model. This is expected since this duration does not cover even a full QBO cycle. Alternatively, we can select 72 weeks, spaced a month apart, covering nearly three QBO cycles, while still being an 18-month long dataset. Offline training the CNN using this \emph{strategically-sampled small-data} regime, denoted as CNN-S3D, yields notably improved results compared to the CNN-SD (Figure S3 in Supporting Information S1). The \emph{a-priori} performance sees the $R^2$ value rise from $0.9$ to $0.98$, but more importantly, at the tails of the GWD PDF, a significant enhancement from $0.14$ to $0.7$. This results in an \emph{a-posteriori} accurate QBO.

%Notably, CNN-S3D's GWD PDF surpasses CNN-EKI's, though CNN-EKI' zonal wind PDF is slightly closer to the true PDF.

This experiment highlights the importance of strategic sampling. Rather than continuous runs, commonly seen in current high-resolution modeling efforts (e.g., \citeauthor{satoh2019global}, 2019; \citeauthor{wedi2020baseline}, 2020), a more effective approach might be to create a library of shorter runs, sampling different regimes/phases important for a given physical process. These runs would provide diverse sampling of various climate and weather conditions without additional computational cost, an approach echoed in \citeA{shen2022library}, and \citeA{sun2023quantifying}.

% without compromising on performance

The results presented in this study provide a proof-of-concept for ``offline-online" learning approach based on the parameterization of GWs in the computationally affordable 1D-QBO model. However, learning from time-averaged statistics necessitates long model simulations during training. Moreover, in complex and high-dimensional parameter spaces, EKI and similar optimization methods require large ensembles and more iterations to achieve an accurate estimate of the optimal parameters. Collectively, these factors can increase the computational cost, and when the forward model is expensive, as is the case for GCMs, the overall cost of EKI can become unfeasible (while here we focus on EKI, it should be noted that other online learning methods such as reinforcement learning suffer from the same challenges). Therefore, efficient strategies are needed to reduce the ensemble size and iterations. For EKI, techniques such as localization, inflation, and regularization are proposed in other studies for these situations \cite{lee2021sampling, tong2022localization, huang2022efficient, iglesias2015iterative, iglesias2016regularizing, iglesias2021adaptive}. 

A common guideline for EKI suggests starting with ensemble numbers that are 10 times the count of parameters. However, a notable observation from this study is that the number of required ensemble members for EKI algorithm to converge does not directly correlate with the number of parameters of the CNN. In our experiments with CNNs containing approximately 1000, 5000, and 10000 parameters, we consistently needed only 200 ensemble members for successful EKI convergence. A simple, yet untested, hypothesis is that increasing the number of parameters might not necessarily expand the dimensions of the parameter space, given the over-parameterized nature of NNs networks. This finding suggests the possibility of using a manageable ensemble size for online training of deep NNs using EKI, when coupled to GCMs.

% While delving deeper into this observation is beyond this paper's scope, 

As discussed earlier, the choice of priors significantly influences EKI's performance. Good priors can notably reduce the number of iterations and potentially the ensemble size. Conversely, poor priors can result in unsuccessful learning. Particularly, we were unable to online train a CNN from a random initialization of its weights and biases: over 95\% of the ensemble members fail at each iteration, preventing the EKI algorithm from converging. However, the SpArK framework revealed that the learned kernels are essentially low-, high-, and band-pass filters. This insight suggest that in the future, instead of using random parameters, we can initialize the weights based on these filter types. Such physics-informed initialization might enhance the effectiveness of EKI training. Exploring this approach is a promising area for future research.

\section{Conclusion}

The process of calibration, upon which the ``offline-online" learning strategy is suggested, is an essential element in the simulation of complex systems and is central to climate model development \cite{balaji2022general}. In this study, we primarily focus on the necessity of online re-training in the context of the \emph{small-data} regime. However, it is essential to highlight that online re-training is probably indispensable when incorporating any data-driven parameterization scheme into the numerical models. This necessity arises from the fundamental differences (e.g., in numerics) between base models, like high-resolution simulations that supply the training data, and the target models, such as operational GCMs. Additionally, potential misalignments between \emph{a-priori} metrics and \emph{a-posteriori} performances further emphasize this need. While our focus in this study is on GWP, the findings can also be applied to other subgrid-scale modeling efforts.

\section*{Open Research}
% AGU requires an Availability Statement for the underlying data needed to understand, evaluate, and build upon the reported research at the time of peer review and publication.

We use the open source software EnsembleKalmanProcesses.jl \cite{oliver2022ensemblekalmanprocesses} for EKI analysis and the \emph{qbo1d} code for the 1D-QBO model simulations, accessible at https://github.com/DataWaveProject/qbo1d.git. 

% Authors should include an Availability Statement for the software that has a significant impact on the research. Details and templates are in the Availability Statement section of the Data and Software for Authors Guidance: \url{https://www.agu.org/Publish-with-AGU/Publish/Author-Resources/Data-and-Software-for-Authors#availability}

% It is important to cite individual datasets in this section and, and they must be included in your bibliography. Please use the type field in your bibtex file to specify the type of data cited. Some options include Dataset, Software, Collection, ComputationalNotebook. Ex: 
% \\
% \begin{verbatim}

% @misc{https://doi.org/10.7283/633e-1497,
%   doi = {10.7283/633E-1497},
%   url = {https://www.unavco.org/data/doi/10.7283/633E-1497},
%   author = {de Zeeuw-van Dalfsen, Elske and Sleeman, Reinoud},
%   title = {KNMI Dutch Antilles GPS Network - SAB1-St_Johns_Saba_NA P.S.},
%   publisher = {UNAVCO, Inc.},
%   year = {2019},
%   type = {dataset}
% }

% \end{verbatim}

% For physical samples, use the IGSN persistent identifier, see the International Geo Sample Numbers section:
% \url{https://www.agu.org/Publish-with-AGU/Publish/Author-Resources/Data-and-Software-for-Authors#IGSN}
%%%%%%%%%%%%%%%%%%%%%%%%%%%%%%%%%%%%%%%%%%%%%%%

\acknowledgments

We are grateful to Oliver Dunbar, Tapio Schneider, Ofer Shamir, and Y. Qiang Sun for valuable discussions. We appreciate the CliMA project for providing public access to EKI. This work was supported by grants from the NSF OAC CSSI program (2005123 and 2004512), and by the generosity of Eric and Wendy Schmidt by recommendation of the Schmidt Futures program (to P.H. and J.A.), by an Office of Naval Research (ONR) Young Investigator Award N00014-20-1-2722 (to P.H.), and by a Rice Academy Postdoctoral Fellowship (to H.P.). Computational resources were provided by NCAR's CISL (allocation URIC0009), and NSF XSEDE (allocation ATM170020).

\end{document}

%% ------------------------------------------------------------------------ %%
%
%  TITLE
%
%% ------------------------------------------------------------------------ %%

%\includegraphics{agu_pubart-white_reduced.eps}

\title{Supporting Information for ``Explainable Offline-Online Training of Neural Networks for Parameterizations: A 1D Gravity Wave-QBO Testbed in the Small-data Regime"}
%
% e.g., \title{Supporting Information for "Terrestrial ring current:
% Origin, formation, and decay $\alpha\beta\Gamma\Delta$"}
%
%DOI: 10.1002/%insert paper number here%

%% ------------------------------------------------------------------------ %%
%
%  AUTHORS AND AFFILIATIONS
%
%% ------------------------------------------------------------------------ %%

% List authors by first name or initial followed by last name and
% separated by commas. Use \affil{} to number affiliations, and
% \thanks{} for author notes.
% Additional author notes should be indicated with \thanks{} (for
% example, for current addresses).

% Example: \authors{A. B. Author\affil{1}\thanks{Current address, Antartica}, B. C. Author\affil{2,3}, and D. E.
% Author\affil{3,4}\thanks{Also funded by Monsanto.}}

\authors{Hamid A. Pahlavan\affil{1}, Pedram Hassanzadeh\affil{1}, M. Joan Alexander\affil{2}}

% \affiliation{1}{First Affiliation}
% \affiliation{2}{Second Affiliation}
% \affiliation{3}{Third Affiliation}
% \affiliation{4}{Fourth Affiliation}

\affiliation{1}{Rice University, Houston, TX, USA}
\affiliation{2}{NorthWest Research Associates, Boulder, CO, USA}
%(repeat as many times as is necessary)

%% ------------------------------------------------------------------------ %%
%
%  BEGIN ARTICLE
%
%% ------------------------------------------------------------------------ %%

% The body of the article must start with a \begin{article} command
%
% \end{article} must follow the references section, before the figures
%  and tables.

\begin{article}

%% ------------------------------------------------------------------------ %%
%
%  TEXT
%
%% ------------------------------------------------------------------------ %%

\noindent\textbf{Contents of this file}
%%%Remove or add items as needed%%%
\begin{enumerate}
\item Gravity Wave Drag in the 1D-QBO Model
\item Stochastic Forcing in the 1D-QBO Model
\item Target Statistics for Online Learning 
\item Figures S1 to S3
% \item Tables S1 to Sx
%if Tables are larger than 1 page, upload as separate excel file
\end{enumerate}
% \noindent\textbf{Additional Supporting Information (Files uploaded separately)}
% \begin{enumerate}
% \item Captions for Datasets S1 to Sx
% \item Captions for large Tables S1 to Sx (if larger than 1 page, upload as separate excel file)
% \item Captions for Movies S1 to Sx
% \item Captions for Audio S1 to Sx
% \end{enumerate}

\noindent\textbf{Introduction}

The Supporting Information delves deeper into the 1D-QBO model configuration. It presents an explanation, along with Figure S1, highlighting the effects of incorporating stochastic forcing into the 1D-QBO model. It provides more details on the target statistics for online learning. Also included are illustrations of the online re-training process via ensemble Kalman inversion (EKI) in Figure S2, and the results of the CNN trained in the \emph{strategically-sampled small-data} regime in Figure S3.

\clearpage

%Delete all unused file types below. Copy/paste for multiples of each file type as needed.
\noindent\textbf{Gravity Wave Drag in the 1D-QBO Model}

In this study, the 1D-QBO model is driven by two gravity waves (GWs) of equal and opposite phase speed with vertical group velocity $c_{gz} = k(u-c_n)^2/N$ for discrete wave $n$, with buoyancy frequency $N = 2.16 \times 10^{-2}$ s$^{-1}$, wavenumber $k=2\pi/40 000$ km$^{-1}$, and phase speed $(c_1, c_2) = (-30, +30)$  m s$^{-1}$. The model domain is from $z_L = 17$ km to $z_T = 35$ km. Subscript $L$ refers to the lower boundary, and subscript $T$ to the upper boundary. The vertical resolution is 500 m. Assuming a constant wave dissipation rate $\alpha = 1.2 \times 10^{-6}$ s$^{-1}$, the wave momentum flux as a function of height for a wave with source momentum flux $F_L = 6.325 \times 10^{-3}$ m$^2$ s$^{-2}$ is
\begin{linenomath*}
\begin{equation} \label{eq:S1}
F_n(u, z) = F_L \,  \mathrm{sgn} \,  (c_n) \, \exp \, \biggl\{ - \int_{z_L}^{z}  \frac{\alpha}{c_{gz}} \,dz^{'} \biggl\}
\end{equation}
\end{linenomath*}

The GWD is proportional to the divergence of momentum flux summed across all waves: 
\begin{linenomath*}
\begin{equation} \label{eq:S2}
G(u, z) = -\frac{\rho_L}{\rho_0} \sum_n \frac{\partial F_n}{\partial{z}}
\end{equation}
\end{linenomath*}

\noindent with density $\rho_0=\rho_L$e$^{(z-z_L)/H_\rho}$ with $\rho_L = 0.1$ kg m$^{-1}$, and $H_\rho = 6$ km. The initial wind at the bottom and top boundaries are set to zero, and they will remain zero at the boundaries. The wind profile is easterly above $\sim$30 km and westerly beneath.

\noindent\textbf{Stochastic Forcing in the 1D-QBO Model}

\begin{figure}[t]
\noindent\includegraphics[width=\textwidth]{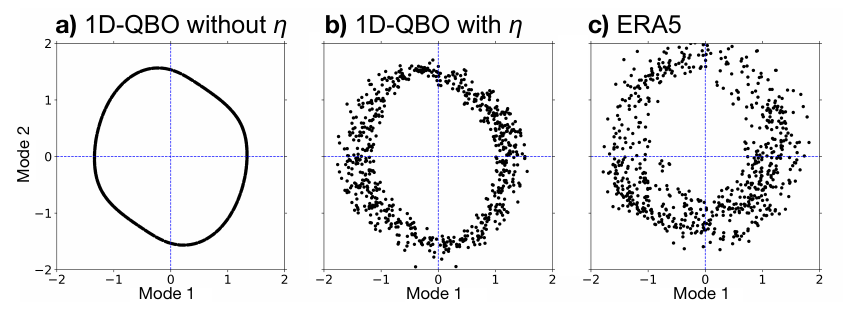}
\caption{The phase diagram of the two leading modes of the empirical orthogonal functions (EOF) for (a) a 1D-QBO model without the stochastic forcing (i.e., the $\eta$ term), (b) a 1D-QBO model with $\eta$, and (c) the QBO as seen in ERA5 reanalysis data.}
\label{Fig. S1}
\end{figure}

The term $\eta$ is a stochastic forcing, randomly generated at each time step from a Gaussian distribution with zero mean and a standard deviation of $4 \times 10^{-6}$. $\eta$ represents the missing physics within the 1D model. Without $\eta$, the model produces a repetitive oscillation with no variability beyond the first QBO cycle. A clear visualization of this repetitive behavior can be observed in the phase diagram constructed based on the two primary modes of the empirical orthogonal functions (EOF) of the oscillation, as shown in Figure~\ref{Fig. S1}a. Introducing stochastic forcing through $\eta$ ensures that each QBO cycle is unique, aligning the model more closely with the observed QBO, as is evident from the phase diagrams shown in Figures~\ref{Fig. S1}b and~\ref{Fig. S1}c.

\noindent\textbf{Target Statistics for Online Learning}

For online learning, we consider a comprehensive list of 85 statistics. This includes:

- The QBO period, evaluated separately for each QBO phase and their average (3 statistics).

- The QBO amplitude, calculated as the standard deviation of the zonal wind at 14 vertical levels (14 statistics).

- Cross-covariance calculations between the levels of 19 and 21.5 km, 18.5 and 29 km, and, 26.5 and 31.5 km, for time lags/leads ranging from -8 to 8 months, which adds up to 51 statistics (3 $\times$ 17).

- The auto-covariance at 34 km for time leads between 1 to 17 months.

Collectively, this amounts to 85 statistics. The derivation of this list involved numerous trial-and-error experiments, combined with a strategic approach to encapsulate the relationship between various QBO levels, to better capture its structure and downward propagation. For instance, the auto-covariance at the 34 km level specifically targets the onset of new QBO phases, which plays a crucial role in determining the QBO period.

\end{article}
\clearpage

% Copy/paste for multiples of each file type as needed.

% enter figures and tables below here: %%%%%%%
%
%
%
%
% EXAMPLE FIGURES
% ---------------
% If you get an error about an unknown bounding box, try specifying the width and height of the figure with the natwidth and natheight options.
% \begin{figure}
% \setfigurenum{S1} %%You can change number for each figure if you want, not required. "S" prepended automatically.
% \noindent\includegraphics[natwidth=800px,natheight=600px]{Figures/Fig.S1.pdf}
% \caption{caption}
% \label{epsfiguresample}
% \end{figure}
%
%
% Giving latex a width will help it to scale the figure properly. A simple trick is to use \textwidth. Try this if large figures run off the side of the page.

\begin{figure}[t]
\noindent\includegraphics[width=\textwidth]{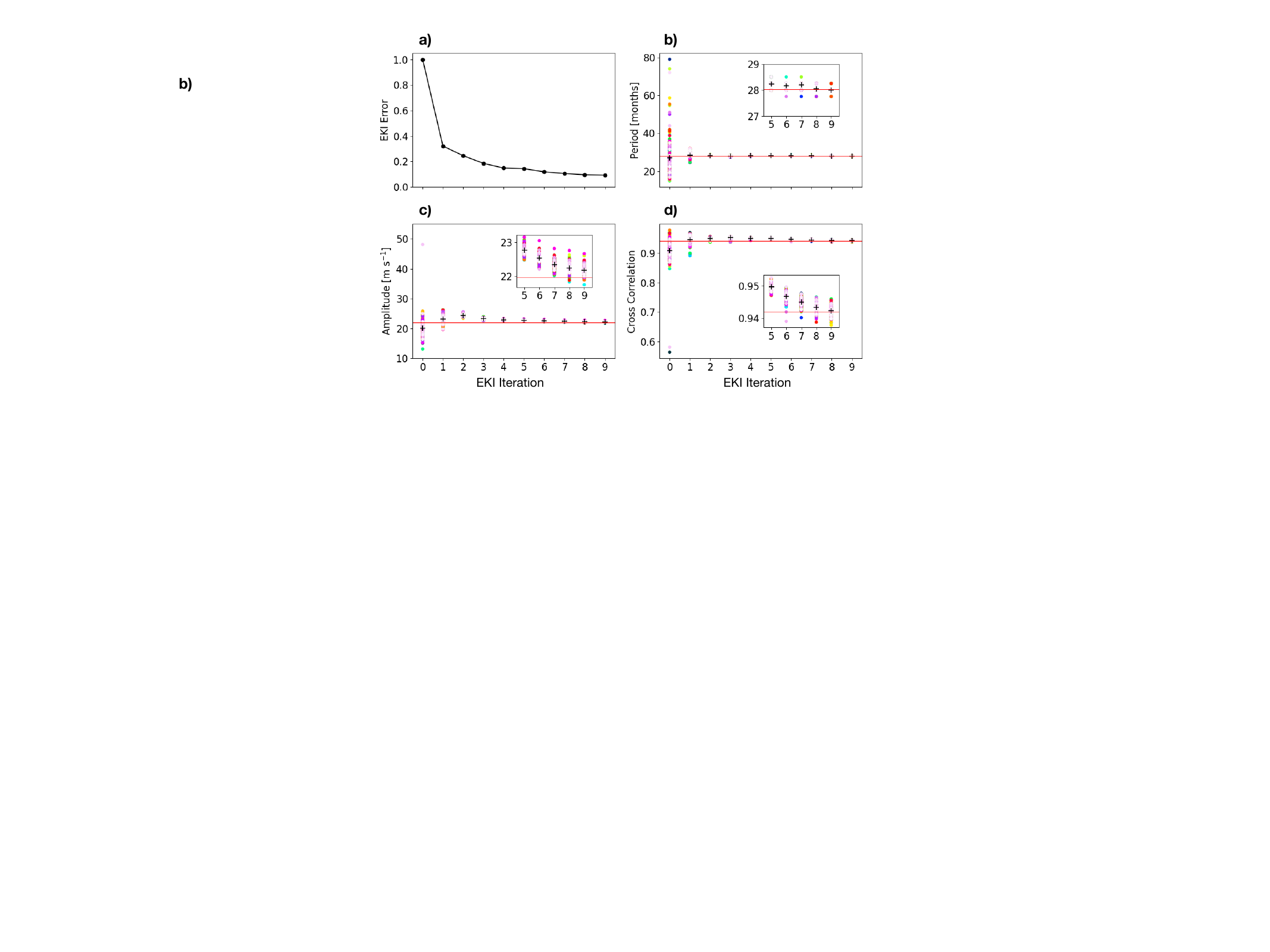}
\caption{ (a) The evolution of EKI error, defined as the average absolute difference between the statistics of the 1D-QBO model with CNN-based GWP, and the true statistics, normalized by the error at the initial iteration. (b) The evolution of the QBO period. Each circle represents an ensemble member, and the black cross indicates the ensemble mean. The inset zooms into the final 5 iterations. (c) As in (b), but for QBO amplitude at an altitude of 30 km. (d) As in (b), but showcasing the cross correlation between the zonal wind at 19.5 km and 22 km with a time-lag of five months. While only three out of the 85 statistics are shown here, a similar pattern is observed across the other targeted statistics.}
\label{Fig. S2}
\end{figure}

\begin{figure}[t]
\noindent\includegraphics[width=\textwidth]{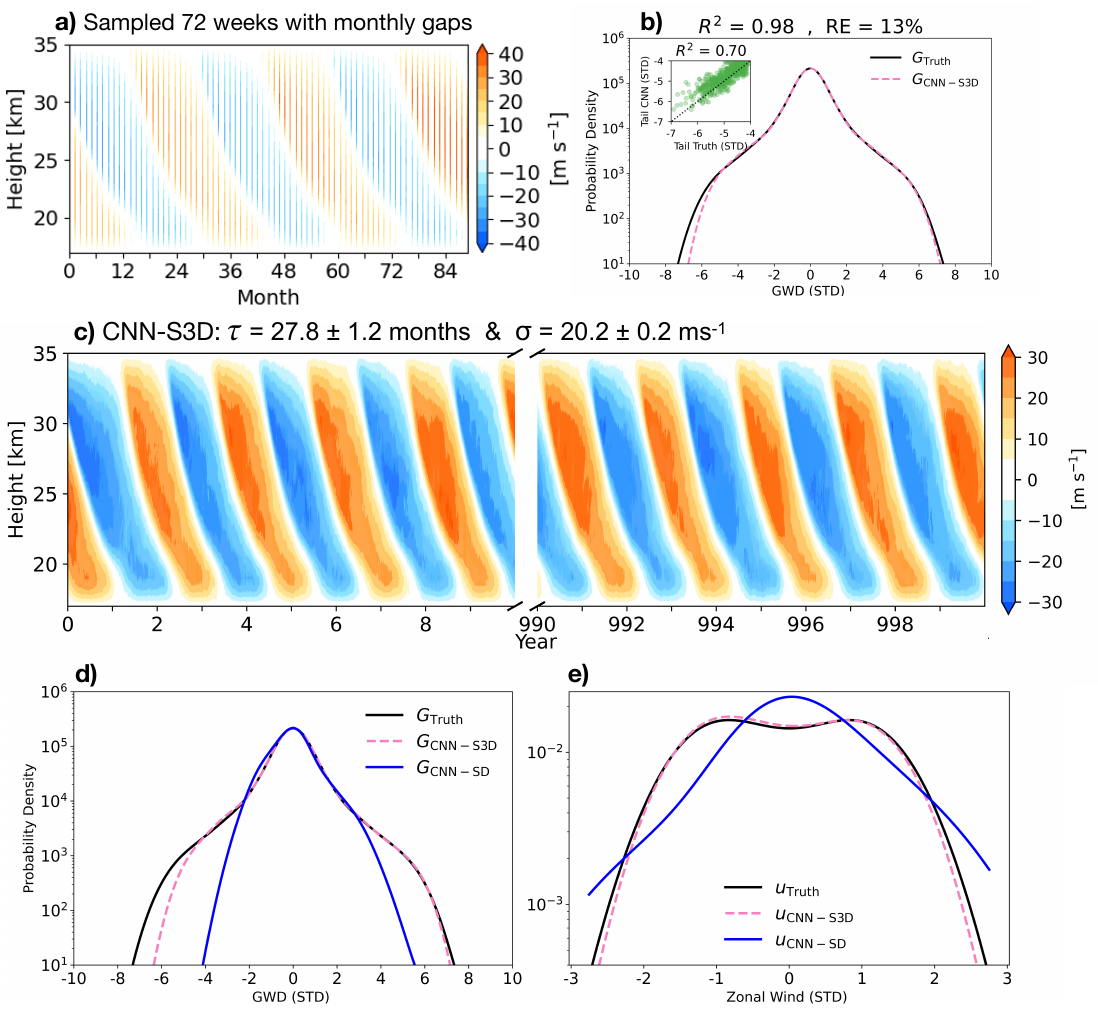}
\caption{ The \emph{a-priori} and \emph{a-posteriori} performances of the CNN trained in a \emph{strategically-sampled small-data} regime, denoted as CNN-S3D. (a) The sampled segments comprise 72 weeks, each separated by a month gap, spanning a total of 90 months. This duration covers nearly three QBO cycles. (b) The \emph{a-priori} performance of the CNN-S3D in predicting GWD. (c) The \emph{a-posteriori} performance of the CNN-S3D. The QBO stays stable for 1000 years, with period and amplitude close to the true QBO. (d) PDFs of the true and \emph{a-posteriori} predicted GWD by CNN-SD and CNN-S3D (G). (e) As in (d), but for zonal wind ($u$).}
\label{Fig. S3}
\end{figure}

%
%
%\begin{figure}
%\noindent\includegraphics[width=\textwidth]{athirdsample.pdf}
%\caption{A pdf test figure}
%\label{pdffiguresample}
%\end{figure}
%
% PDFLatex does not seem to be able to process EPS figures. You may want to try the epstopdf package.
%
%
% ---------------
% EXAMPLE TABLE
%
%\begin{table}
%\settablenum{S1} %%Change number for each table
%\caption{Time of the Transition Between Phase 1 and Phase 2\tablenotemark{a}}
%\centering
%\begin{tabular}{l c}
%\hline
% Run  & Time (min)  \\
%\hline
%  $l1$  & 260   \\
%  $l2$  & 300   \\
%  $l3$  & 340   \\
%  $h1$  & 270   \\
%  $h2$  & 250   \\
%  $h3$  & 380   \\
%  $r1$  & 370   \\
%  $r2$  & 390   \\
%\hline
%\end{tabular}
%\tablenotetext{a}{Footnote text here.}
%\end{table}
% ---------------
%
% EXAMPLE LARGE TABLE (UPLOADED SEPARATELY)
%\begin{table}
%\settablenum{S1} %%Change number for each table
%\caption{Time of the Transition Between Phase 1 and Phase 2\tablenotemark{a}}
%\end{table}